\documentclass[epsfig,12pt]{article}
\usepackage{epsfig}
\usepackage{graphicx}
\usepackage{amsmath}
\usepackage{chngcntr}
\usepackage{wrapfig}
\usepackage{caption}
\usepackage{subcaption}

\newcommand{\beq}{\begin{equation}}   
\newcommand{\eeq}{\end{equation}}
\newcommand{\beqn}{\begin{eqnarray}}   
\newcommand{\eeqn}{\end{eqnarray}}

\newcommand{\gsim}{\lower.7ex\hbox{$
\;\stackrel{\textstyle>}{\sim}\;$}}
\newcommand{\lsim}{\lower.7ex\hbox{$
\;\stackrel{\textstyle<}{\sim}\;$}}

\begin{document}
\begin{flushright}
FTPI-MINN-14/34, UMN-TH-3405/14\\
\end{flushright}

\vspace{0.3cm}

\begin{center}
{\Large\bf Resurgence, Operator Product Expansion, \\[1mm]  and Remarks on Renormalons in \\[4mm] 
Supersymmetric Yang-Mills}

\vspace{5mm}

{\large M. Shifman}

\vspace{0.6cm}
{\em William I. Fine Theoretical Physics Institute, University of Minnesota,
Minneapolis, MN 55455, USA}

\vspace{2cm}

{\bf Abstract}

\end{center}

\vspace{0.2cm}

{\small  
I discuss similarities and differences between the resurgence program in quantum mechanics
and the operator product expansion in strongly coupled Yang-Mills theories.   
In ${\mathcal N}=1$ super-Yang-Mills renormalons possess peculiar features that make them 
different from renormalons in QCD.  Their conventional QCD interpretation does not seem to be applicable
in supersymmetric Yang-Mills in a straightforward manner. 

This is a write-up of my talk at the Conference ``Resurgence and Trans-series in Quantum, Gauge and String Theories,"  June 30 - July 4, 2014, CERN (see http://cds.cern.ch/record/1741028). }

\newpage

\section{Introduction}
\label{int}

The notion of {\em resurgence} and trans-series associated with it -- a breakthrough discovery\,\footnote{For a pedestrian review understandable to physicists (at least, in part) and an exhaustive list of references see \cite{edg,dun}.} in constructive mathematics in the 1980s mostly associated with the name of Jean Ecalle -- gradually spread in mathematical and theoretical physics.  
I was impressed by diverse and numerous applications of these ideas which were discussed at this conference in the excellent talks of J.~Zinn-Justin, M. Berry, U. Jentschura, G. Dunne, M. Beneke, and others. The topic closest to my talk is resurgence in quantum mechanics. We could see that in quantum mechanics it works well, and trans-series 
of the type
\begin{eqnarray}
E(g^2)&=&E_{\rm PT, \,\,regularized}(g^2) 
\nonumber\\[3mm]
&+&  \sum_{k=1}^\infty \sum_{l}\sum_{p=0}^\infty 
\underbrace{ \left(\frac{1}{g^{2N+1}}\,\exp\left[-\frac{c} {g^2}\right]\right)^k }_{ \rm k-instanton}\,
{ \left(\log \frac{c}{g^2}\right)^l}\, \underbrace{c_{k, l, p}  g^{2p}}_{\rm regularized\,\,  PT}
\label{trans}
\end{eqnarray}
can be derived for all energy eigenvalues ($g^2$ is assumed to be small; the subscript PT stands for perturbation theory).

 In weakly coupled field theories trans-series could be perhaps constructed, although conclusive arguments have not  yet been presented. One of the tasks which I formulated for myself  is to explain why resurgence -- being conceptually close to the operator product expansion (OPE) does {\em  not } work in strongly coupled field theories,  for instance, quantum chromodynamics, (QCD). It is worth noting that OPE existed in QCD from mid-1970s, and in its general form the late 1960s.
It grew from a  formalism  which had been suggested by K. Wilson before the the advent of QCD. 
The first part of my talk will be devoted to this issue.

In the second part I will focus on a more technical aspect:  peculiarities of the factorial divergence of perturbation theory
in ${\mathcal N}=1$ super-Yang-Mills (SYM). Do far renormalons in SYM were  scarcely discussed.  We (I mean M. \"Unsal, G. Dunne, A. Cherman, and myself) exchanged a few remarks on this subject almost a year ago, when we all participated in one and the same conference and shortly after. Later I returned to this issue but no final conclusion has been reached. To a large extent this question remains open. 

\section{The simplest quantum-mechanical examples }
\label{simplest}

\subsection{Anharmonic  oscillator}
\label{ano}

\begin{figure}[h]
   \epsfxsize=5cm
   \centerline{\epsffile{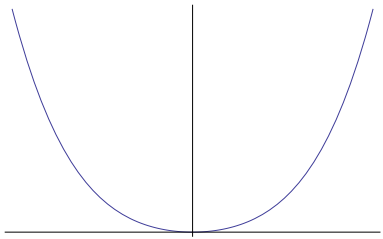}}
\caption{\small$V(x)$ in the anharmonic oscillator problem (\ref{1}).
\label{pl1}}
\end{figure}

Let us consider one-dimensional anharmonic oscillator,
\beq
{\mathcal H} = \frac12 p^2 + \frac12\omega^2 x^2 +g^2x^4\,,
\label{1}
\eeq

see Fig. \ref{pl1}.
For definiteness we will focus on the ground state energy $E_0$.
There exists a well-defined procedure of constructing $E_0$ order by order in perturbation theory, to any finite order,
\beq
E_0 =  \frac{\omega}{2} \left(1 + c_1 g^2 + c_2 g^4 +....
\right).
\label{2}
\eeq
Nevertheless, Eq. (\ref{2}) does not define the ground state energy.
Indeed, the coefficients $c_k$ are factorially divergent at large $k$ \cite{ZJ},
\beq
c_k \sim (-1)^k B^{-k} k!\,,\qquad k\gg 1\,,
\label{3}
\eeq
where $B= \frac13 \omega^3$ is the so-called bounce action.\footnote{Equation (\ref{3}) is slightly simplified.
For a more precise formula see \cite{ZJ}.} Thus, the sum in (\ref{2}) needs a regularization.

\begin{figure}[h]
   \epsfxsize=5cm
   \centerline{\epsffile{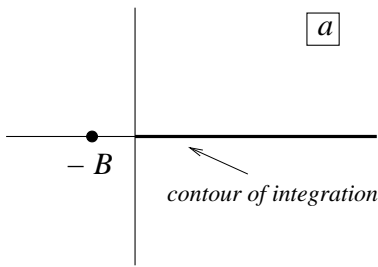}}
\caption{\small The perturbative series in the anharmonic oscillator problem is Borel-summable. The $g^2$ series for $E_0$ is sign alternating;
$f(a)$ has a singularity on the real negative semi-axis in the Borel parameter complex plane. $a$ is the Borel parameter.
\label{wd}}
\end{figure}

In the simplest case under consideration an appropriate (and exhaustive) regularization is provided by the Borel transformation ${\mathcal B}$,
\beq
{\mathcal B}E_0\equiv \frac{\omega}{2} \left(1 + \sum_{k=1}^\infty \frac{1}{k!}c_k g^{2k}\right) \equiv \frac{\omega}{2}\,
f(g^2)
\,.
\label{4}
\eeq
The Borel transformation introduces $1/k!$ in each term of the series (\ref{2}) rendering it convergent.
Moreover, if the convergent series   
\beq
1 + \sum_{k=1}^\infty \frac{1}{k!}c_k g^{2k} \equiv f(g^2)
\eeq
 which defines the Borel function $f(g^2)$ has no singularities on the real  positive semi-axis $g^2\geq 0$ then
one can  obtain the ground-state energy $E_0$ starting from the well-defined expression for ${\mathcal B}E_0$ and using the Laplace transformation,
\beq
E_0 = L\left({\mathcal B}E_0\right) \equiv \frac{\omega}{2}\,\int_0^\infty \, da\, g^{-2} \exp\left(- \frac{a}{g^2}\right)f(a)\,.
\label{6p}
\eeq
This procedure is usually referred to as the Borel summation. Thus, the perturbative expansion in the anharmonic oscillator is Borel-summable due to the fact that the singularities of $f(a)$ are on the {\em negative} real semi-axis. 
Indeed, assume that $f(a)$ has a pole at $a=-B$ (see Fig. \ref{wd}), namely,
\beq
f(a) = \frac{B}{a+B}\,,\qquad B= \frac{\omega^3}{3}\,.
\label{7p}
\eeq
Then the integral (\ref{6p}) is well-defined.
At the same time, expanding (\ref{7p}),
\beq
f(a) =\sum_{k=o}^\infty \,(-1)^k \left(\frac{a}{B}
\right)^k\,,
\label{8p}
\eeq
and substituting this series in (\ref{6p}) we immediately arrive at (\ref{3}). 

The fact that the position of the singularity in the $a$ plane is to the left of the origin  and   that the series is sign-alternating are in one-to-one correspondence with each other.

\begin{figure}[t]
   \epsfxsize=5cm
   \centerline{\epsffile{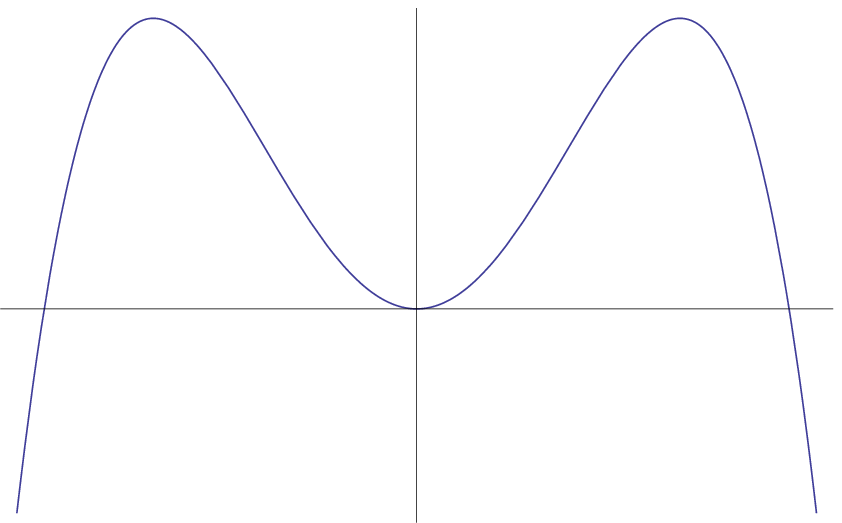}}
\caption{\small The same potential with the replacement $g^2\to -g^2$, to be denoted as $\tilde V(x)$.
\label{pl2}}
\end{figure}

Exactly fifty years ago Vainshtein identified \cite{va} the physical meaning of the factorial growth
of the coefficients (\ref{3}) and explained why the underlying singularity in the Borel parameter plane is on the negative semi-axis.
Changing the sign of $g^2$
from positive to negative, $g^2 \to - g^2$, one converts a stable potential $V(x)$ in (\ref{1}) into an unstable potential 
$\tilde V(x)$ presented in Fig \ref{pl2}, allowing for the leakage of the wave function to large distances.

\begin{figure}[h]
\centering
\includegraphics[totalheight=0.16\textheight, angle=180]{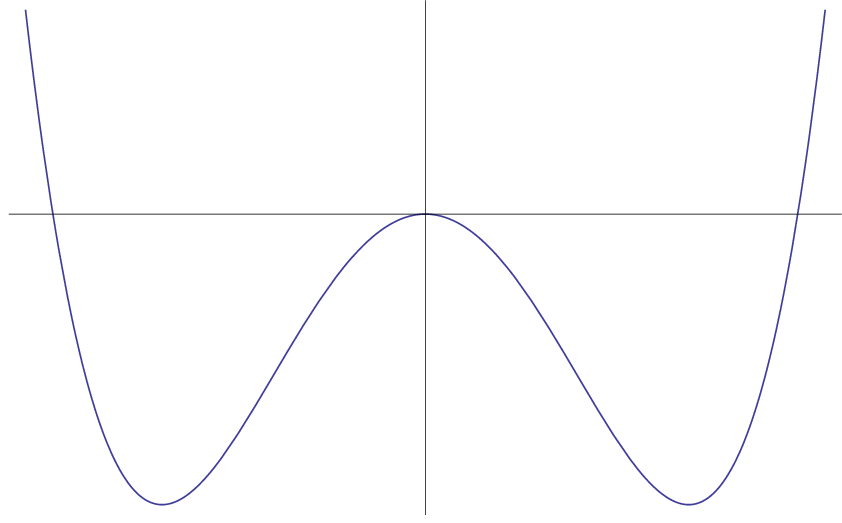}
\caption{\small An effective potential in Euclidean time. This potential presents a sign reflection of that in Fig.
\ref{pl2}, i.e. is  $-\tilde V(x)$. It vanishes at $x=0$  and at $x=\pm x_*$ where  $x_* = \frac{\omega}{\sqrt{2}\,g}$.}
\label{minus}
\end{figure}

In the leaking potential $\tilde{V}$ the energy to the $0$-th eigenvalue acquires an imaginary part (as well as other energy eigenvalues). This imaginary part can be easily determined. Indeed, after the Euclidean time rotation effectively the potential $\tilde V(x)\to - \tilde V(x)$, as shown in  Fig. \ref{minus}. Then 
the so-called bounce trajectory becomes classically accessible.\footnote{See e.g. Chapter 7 in \cite{ms}.}  The bounce trajectory starts at $x=0$, slides to the right, bounces off at $x_* = \frac{\omega}{\sqrt{2}\,g}$ and then returns to  the point $x=0$.  The Euclidean action on the bounce trajectory is readily calculable,
\beq
A_{\rm bounce} = \frac{B}{g^2}\,,
\label{9p}
\eeq
 where $B$ is defined in Eq. (\ref{7p}). In this way we obtain that 
\beq
{\rm Im} \,E_0 =  \frac{\pi\,\omega}{2}\,\frac{B}{g^2}\, \exp\left(-\frac{B}{g^2}
\right).
\label{9pp}
\eeq
Now one can calculate  the ground state energy for the original potential in Fig. \ref{pl1} by using (\ref{9pp}) and a dispersion relation in the coupling constant \cite{va},
\begin{eqnarray}
E_0 &=& \frac{1}{\pi}\, \int_0^\infty d\tilde{g}^2\,\frac{1}{g^2 +\tilde{g}^2}\,{\rm Im}\, E_0\left(\tilde{g}^2\right)
\nonumber\\[3mm]
&=& \frac{\omega}{2} \,\int \,dz \,\frac{1}{1 + \frac{g^2}{B} \, z} \,e^{-z}\,.
\end{eqnarray}
The last expression reproduces the  series in (\ref{2}) and  (\ref{3}) with its sign alternation and factorial divergence of the coefficients. Both features are explained by the imaginary part in (\ref{9pp}) proportional to $\exp (-B/g^2)$.

Summarizing, the perturbative expansion for the anharmonic oscillator is factorially divergent; however, the Borel summability allows one to find the closed, well-defined and exact expressions for the energy eigenvalues.  The physical meaning of the factorial divergence, as well as the sign alternation, are fully understood. Now we will pass  to a more complicated but more interesting non-Borel  summable case.

\subsection{Double-well potential}
\label{dowe}

The double-well problem is described by the Hamiltonian
\beq
{\mathcal H} = \frac12 p^2 - \frac14\omega^2 x^2 +g^2x^4\,,
\label{12}
\eeq
i.e. the sign of the $O(x^2)$ term is changed, and the point $x=0$ becomes unstable. Instead, two stable minima develop at $x_*=\pm\,x_*$ where 
$$x_*=
 \omega/2\sqrt{2}g\,.$$ The shape of the double well potential is depicted in Fig. \ref{minus}.
Classically each of the two minima  $x=\pm\, \omega/2\sqrt{2}g$ present stable solutions of the system at hand. 
Quantum-mechanically zero point oscillations about the minima occur. Taking into account anharmonicity 
near the minima, we generate a perturbative series for the ground state energy. This is in perturbation theory.
In fact, the two minima are connected by the tunneling trajectory (instanton) in Euclidean time. The instanton action is
\beq
S_{\rm inst} = \frac{\omega^3}{12g^2}\,,
\eeq
see e.g. \cite{mi2}.  In  what follows it will be convenient to introduce
\beq
B_{\rm inst} =  g^2 S_{\rm inst} =\frac{\omega^3}{12}\,.
\eeq

\begin{figure}[h]
   \epsfxsize=5cm
   \centerline{\epsffile{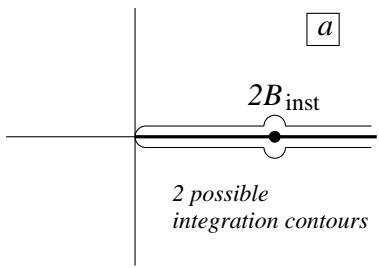}}
\caption{\small The $g^2$ series in the double well problem  is not sign alternating;
$f(a)$ has a singularity on the real negative semi-axis in the Borel parameter complex plane at  $a=2B_{\rm inst} $ 
where $a$ is the Borel parameter.
\label{ianbs}}
\end{figure}

At small $g^2$ the ground state energy is close to $\omega/2$ plus corrections in $g^2$ and nonperturbative corrections of the type $\exp(-c/g^2)$. A crucial distinction from the anharmonic oscillator discussed in Sec. \ref{ano} is the fact the the $g^2$ series in this case will not be sign-alternating (although still factorially divergent), corresponding to a singularity in the Borel function at a real positive value of $a=2B_{\rm inst}$, i.e.  on the integration contour, see
Fig. \ref{ianbs}
Thus, one has to rethink the  Borel summation procedure.

Equation (\ref{6p}) is replaced by
\beq
E_0 = L\left({\mathcal B}E_0\right) \equiv \frac{\omega}{2}\,\int_0^\infty \, da\, g^{-2} \exp\left(- \frac{a}{g^2}\right)f(a)\,,
\label{6pp}
\eeq
where roughly speaking
\beq
f(a) = \frac{-2B_{\rm inst}}{a-2 B_{\rm inst}}\,.
\label{7pp}
\eeq
Then, instead of Eq. (\ref{3}) we obtain
\beq
c_k = k!\, \left(2B_{\rm inst}
\right)^{-k}\,.
\eeq
The perturbative series is not sign alternating, unlike the case of the anharmonic oscillator.

Let us pause here to have a closer look at the above results.
In fact, the integral (\ref{6pp}) is undefined: the integration along the real positive semi-axis cannot be performed since we hit a singularity. It must be circumvented either along the upper or lower small semicircles as shown in Fig.  \ref{ianbs}. Depending on whether we choose the upper or lower semicircle we get an {\em imaginary} contribution
\beq
\left(\Delta E_0\right)_{\rm Borel} =\pm \pi\,i \left(-\frac{2B_{\rm inst}}{g^2}\right) \exp\left(-\frac{2B_{\rm inst}}{g^2}\right)\,.
\label{18p}
\eeq

However, in the case of the double well potential the system is stable and does not decay, implying that
the ground state energy must be strictly real. $\left(\Delta E_0\right)_{\rm Borel}$ must be canceled by something, and, indeed, it is canceled by a contribution coming from the instanton-anti-instanton (IA) pair. The position of singularity at $2B_{\rm inst}$ in Fig. \ref{ianbs} prompts us that it is a pair of instantons which is important.

The IA pair is only an approximate saddle point. There is an attraction potential which is very shallow when they are far apart. As usual, approximate saddle points require a regularization. One of regularizations which is very helpful at least for qualitative purposes is considering the IA pair at a finite (rather than zero) energy $E$, along the lines described e.g. in \cite{ms}, Secs. 23.2 and 23.3.  Then the imaginary part of the IA contribution
reduces to $\exp (-2S_{\rm inst})$
with a known pre-exponent, and cancels the imaginary part in (\ref{18p}).
The real part of the IA contribution is proportional to $\omega T_* \sim \log \left(S_{\rm inst}\right)(\omega/E)$
where $T_*$ is the critical $IA$ separation and the value of $E$ relevant to the problem is $E\sim \omega$. 
Thus, the real part of the IA contribution reduces to $(\log S_{\rm inst})\times \exp (-2S_{\rm inst})$ times a known power of 
$S_{\rm inst}$
in the pre-exponent. For a more careful calculation see \cite{Bogomolny,Zinn,Yung}.

If we write the ground-state energy in the form
\beqn
E_0 &=& \frac{\omega}{2}\,\,{\rm P}\! \int_0^\infty \, da\, g^{-2} \exp\left(- \frac{a}{g^2}\right)f(a) 
\nonumber\\[3mm]
&+& (S_{\rm inst})^p (\log S_{\rm inst}) \exp (-2S_{\rm inst}) \,,
\label{19pp}
\eeqn
where P stands for the principle value of the integral this expression is well-defined and, being expanded, generates the perturbative series in its entirety.\footnote{Equation (\ref{19pp}) is to be compared with the general trans-series formula (\ref{trans}).} Strictly speaking the second line is oversimplified, since the pre-exponent in the second line will also be represented by an infinite $g^2$ series with factorially divergent coefficients. To amend this series we will have to include 2I-2A contribution, and so on. Let us refer to this formula as the minimal Borel procedure (MBP). The MBP formula contains all information one can squeeze from perturbation theory. It still lacks something. In order to understand what this something is let us make a digression.

As well-known, perturbation theory (PT) describes fluctuations of a quantal system around
classical minima of the potential. 
In the case at hand we have two degenerate minima reflecting a $Z_2$ symmetry of the potential. Let us choose one of them for the ``unperturbed" Hamiltonian, for instance,
\beq
H_0= \frac{p^2}{2} + \frac{\omega^2}{2} (x-x_*)^2\,.
\label{hao}
\eeq
All cubic and quartic terms from the expansion of the potential (\ref{12}) are referred to $ H_{\rm int}$.
Perturbation theory in $H_{\rm int}$ is well defined in any order.

$H_0$ does not  know about the second vacuum vacuum, but 
high-order corrections will reflect the existence of the second vacuum indirectly, through factorial divergence of the 
PT series. Perturbation theory in $H_{\rm int}$ requires only the knowledge of the
unperturbed eigenfunctions and eigenvalues (i.e those of the harmonic oscillator (\ref{hao})). The eigenfunctions of $H_0$ should be square-normalizable, and  no other requirements are imposed.

Next, we define the sum of the factorially divergent series  as MBP plus IA.
Using this procedure we would conclude that the system at hand has two degenerate ground states: the $Z_2$  restoration in the vacuum is still absent.

This fact -- restoration of $Z_2$ -- does not ensue with necessity from the amended PT series. 
It presents an additional information on the global vacuum structure: a $Z_2$ order parameter drastically changes compared to its PT value, accordingly the degeneracy of the ground state is lifted.
This effect is proportional to $ \exp(-S_{\rm inst})$, as opposed to 
$\exp(-2S_{\rm inst})$ reflecting the corresponding singularity in the Borel plane at $2B_{\rm inst}$.\footnote{
The instanton can leak to another minimum and then anti-instanton will return the system back to the original one. That is the origin of  the  
$\exp(-2S_{\rm inst})$ factor. The splitting between the ground state and the first excitation is due to a single instanton
which connects two ``prevacua." This effect is proportional to $\exp(-S_{\rm inst})$.}

Conceptually, this  is similar to the chiral symmetry breaking in the chiral limit in QCD.
No matter what one does with PT, one  will not see any splitting between the axial and vector quark two-point functions.  One  has to infer the global vacuum structure of QCD other sources.

\section{Asymptotically free field theories}

We arrived at a point where it would be natural to pass from quantum mechanics to asymptotically free field theories.
Up to a certain point we will be able to proceed along the lines outlined in Sec. \ref{simplest}. There are two cases
allowing one to go all the way up to complete resurgence: (a) if a given field theory is exactly solvable (in which case this is a triviality), or (b) if it is weakly coupled (perhaps, after a certain deformation) and hence can  be treated quasiclassically. In the latter case complications that arise are of a technical nature. Today we are aware of quite a few examples of this type which have been identified and studied in the past.

However, the most interesting theories are QCD and its relatives. They are special because QCD is {\em the} theory of Nature, describing quark-gluon dynamics. They are strongly coupled in the infrared domain (IR) where it is impossible to treat them quasiclassically -- perturbation theory fails even qualitatively. It does not capture drastic rearrangement of the vacuum structure related to confinement. 

I would like to discuss the following question: how far can we go in the resurgence program in these theories? We will see that a certain procedure suggested in the late 1960s \cite{wi}  and implemented in QCD in the 1970s \cite{svz}  allows one to advance rather far, although, unfortunately, not to the very end. This is as good as it gets...

The Lagrangian of QCD has the form (in the chiral limit)
\beq
{\mathcal L} = -\frac{1}{4g^2} G_{\mu\nu}^a G^{\mu\nu\, a} + \sum \bar\psi i\, / \!\!\! \! D\psi\,. 
\label{qcdl}
\eeq
where the sum runs over the massless quark flavors, and $\psi$ is the quark field in the fundamental representation 
of SU$(N)$. In actual world $N=3$, but in theoretical laboratory one is free to consider any value of $N$. If we drop the quark term we are left with the 
$G_{\mu\nu}^2$ gluon term. This is pure Yang-Mills theory. Moreover, $g^2$ in front of the gluon term is the asymptotically free gauge coupling.

As we know, this is a strongly coupled theory. The Lagrangian is defined at short distances  in terms of gluons and
quarks, while at large distances of the order of $\gsim \Lambda_{\rm QCD}^{-1}$  we deal with hadrons, e.g.  pions and protons. Certainly, the latter are connected with quarks and gluons in a divine way, but this connection is highly nonlinear, non-local and is not amenable to analytical description at the moment. Moreover, the very existence of massless pions and massive protons is due to a dramatic restructuring of the QCD vacuum reflecting spontaneous breaking of the chiral symmetry. The latter phenomenon is only possible at very strong coupling. Perhaps, in the future string theory will be able to provide an adequate description, but as they say ``the future is not ours to see..."

Another -- a much simpler -- example is two-dimensional CP$(N-1)$ model with a varying degree of supersymmetry (or no supersymmetry at all). The action of the model is
\beq
S =\int d^2x\,\left(\sum_{i,\bar j =1}^{N-1}
G_{i\bar j}\,\partial_{\mu}\phi^{\dagger \bar j}\, \partial^{\mu}\phi^{i} + {\rm fermions}\right)
\eeq
where
\beq
G_{i\bar j}=\frac{2}{g^{2}}\Bigg(\frac{\delta_{i\bar j}}{\chi}-\frac{\phi^{\dagger\,i}\phi^{\bar j}}{\chi^{2}}\Bigg)\,,\qquad
\chi =1+\sum_{m}^{N-1}\phi^{\dagger\, m}\phi^{m}\,,
\eeq
and $g^2$ is the asymptotically free coupling constant. 
In the large-$N$ limit this model is exactly solvable \cite{A,B,C}. To the leading order in $1/N$ the solution is known but cannot be expressed in terms of (\ref{trans}) because instantons are irrelevant at strong coupling. Since the solution is known, one can still present it in the form of a generic trans-series. In the first subleading $1/N$ correction we return to a generic contrived situation, to be discussed below, similar to that in QCD.

A common feature of both theories above as well as many others from this class, is the fact that the coupling constant 
is not a {\em bone fide} constant; it runs. In more detail
\beq
g^2(Q) = \frac{B_{\rm inst} \rule{0mm}{6mm}}{\beta_0 \log (Q/\Lambda)}
\eeq
where
\beq
B_{\rm inst} = \left\{ \begin{array}{l}
8\pi^2 ,\,\,\, {\rm QCD }\\[1mm]  4\pi ,\,\,\,\,\, {\rm CP}(N-1)
\end{array}\right., \quad 
\beta_0= \left\{ \begin{array}{l}
\frac {11}{3} N ,\,\,\, {\rm YM}\\[2mm]  N ,\,\,\,\,\,\,\,\, {\rm CP}(N-1)
\end{array}\right. .
\eeq
and $Q$ is an appropriate momentum scale (assuming $Q\gg \Lambda$.)
Here $\beta_0$ is the first coefficient of the $\beta$ function. In the upper line on the right it is given for pure Yang-Mills. When characteristic values of $Q$ become close to $\Lambda$, the running constant is undefined and all calculations in terms of gluons and quarks become meaningless. 

As wee see, the genuine parameter of QCD is not dimensionless $g^2$ but, rather the dynamical QCD scale 
$\Lambda$ invisible in the classical Lagrangian. That's the phenomenon of dimensional transmutation inherent to all
strongly coupled asymptotically free field theories. 
The series in $g^2$ becomes the series in $ 1/\log \frac{Q}{\Lambda}$, exponential terms $\exp(-c/g^2(Q)$ reduce to powers 
$$
\left(\frac{\Lambda}{Q}
\right)^{c\beta_0/8\pi^2}\,,
$$
while exponential in $Q$ terms $\sim \exp (-cQ/\Lambda)$, which also appear in QCD and similar theories, in $g^2$ perturbation theory have to emerge from  
$$
\exp (-c\, \exp(\tilde c/g^2(Q)))\,.
$$

Complete failure of quark-gluon calculations at $Q\sim \Lambda$ blocks the program of ``analytic" resurgence in terms of trans-series in QCD. However, some kind of resurgence is possible, known as the operator product expansion.  Now I proceed to a more systematic (albeit brief) discussion of OPE.

\section{OPE vs trans-series}
\label{ots}

Instead of a general introduction to Wilson's operator product expansion which would require a lot of  
time\,\footnote{For a review of OPE in QCD see \cite{ope1}.}
I will briefly discuss OPE from a somewhat nonstandard standpoint: following the logic  of Sec. \ref{simplest}
devoted to resurgence in quantum mechanics. 

\begin{figure}[h]
    \centering
    \begin{subfigure}{.49\textwidth}
        \centering
        \includegraphics[width=5cm]{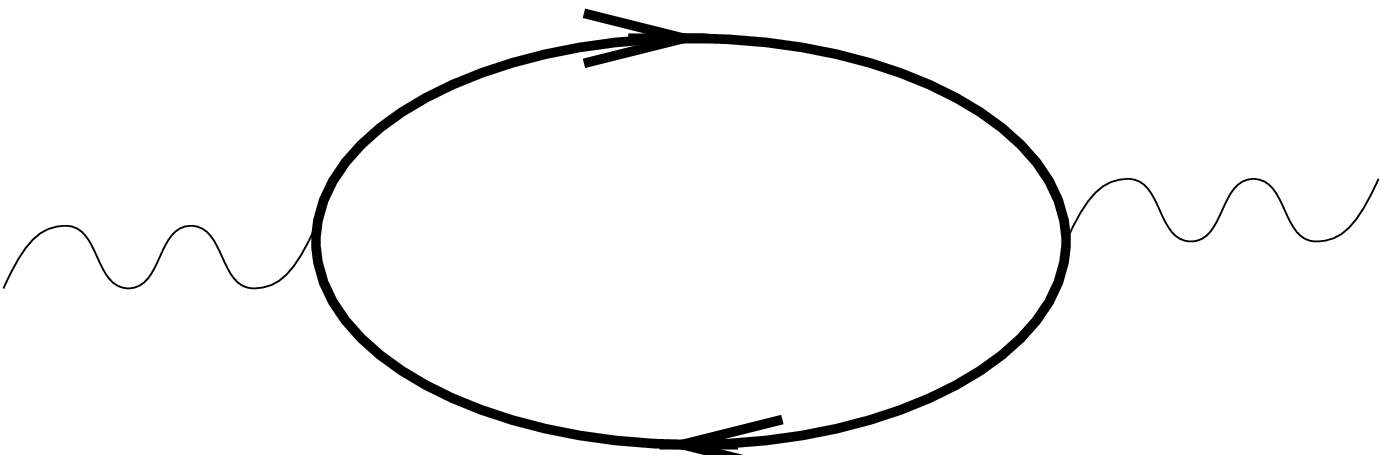}
        \caption{$O(\alpha^0)$}
        \label{fig:t}
    \end{subfigure}
    \begin{subfigure}{.49\textwidth}
        \centering
        \includegraphics[width=5cm]{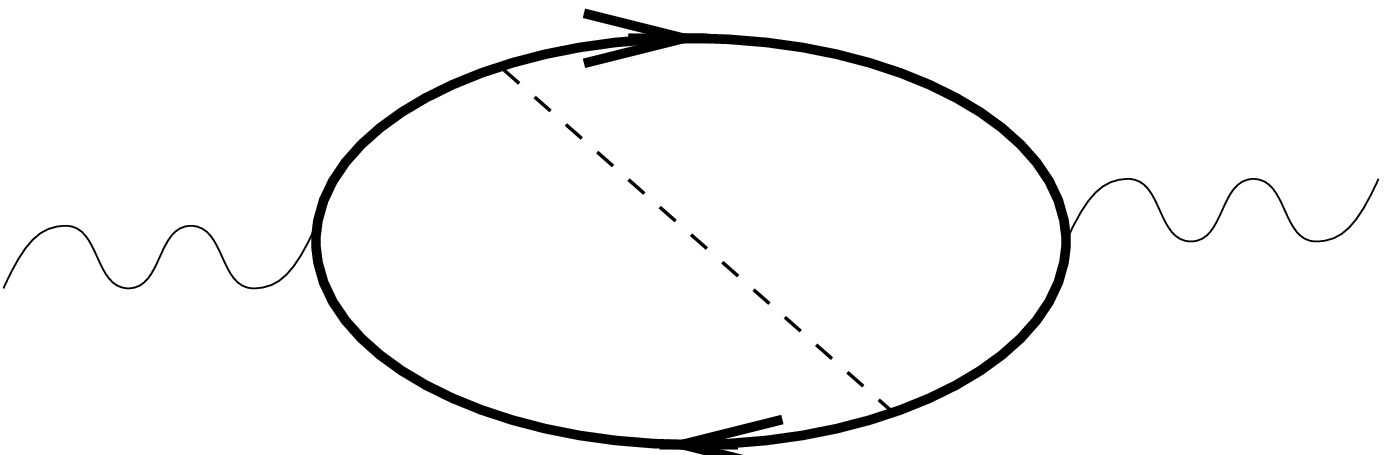}
        \caption{$O(\alpha^1)$}
        \label{fig:w}
    \end{subfigure}
    \caption{\small The leading and the next-to leading terms in the expansion of the Adler function. The external 
    current $j_\mu$ injecting the 
quark-antiquark pair in the vacuum (and then annihilating it) is 
denoted by wavy lines. }
    \label{fig:ef}
\end{figure}

Let us start our discussion from the two-point function
 \begin{eqnarray}
\Pi_{\mu\nu} (q) &=&
i\, \int\,d^4x\,e^{ i q x}\,\left\langle T\,
\left( j_\mu(x) j_\nu(0)  \rule{0mm}{4mm} \right) \rule{0mm}{4mm}\right\rangle 
\nonumber\\[2mm]
&=& \left(q_\mu q_\nu-q^2 g_{\mu\nu}\right)\,\Pi(Q^2)\,,
\label{currentcorr}
\end{eqnarray}
where $
j_\mu = \bar\psi\gamma_\mu\psi$ is the quark current, and we denote
 \beq
 Q^2=-q^2\,,
 \label{eucl}
 \eeq
 so that in the Euclidean domain $Q^2$ is positive. We will limit ourselves to large values of the Euclidean momentum,
  $Q\gg \Lambda$, so that perturbation theory can be used. 
In fact, instead of  $\Pi(Q^2)$, for technical reasons  it will be convenient analyze the so-called Adler function defined as
\begin{equation}
\label{adlerdef}
D(Q^2)=- 4 \pi^2\,Q^2\,\frac{d\Pi(Q^2)}{dQ^2}\,.
\end{equation}
The first two terms in the expansion of the Adler function are defined by the diagrams in Fig. \ref{fig:ef} where
the coupling constant
\beq
\alpha \equiv \frac{g^2}{4\pi}\, .
\eeq
Given the external momentum $Q$ flowing through the wavy line, it is easy to see that it is the running coupling 
$\alpha (Q)$ that enters in Fig. \ref{fig:w}. Indeed, the momentum flowing through the gluon line in Fig. \ref{fig:w})  
is  $k\sim Q$.

\vspace{10mm}

\begin{figure}[h]
   \epsfxsize=7cm
   \centerline{\epsffile{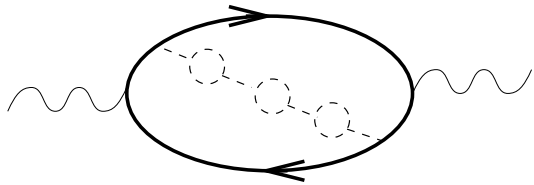}}
\caption{\small The bubble-chain diagrams representing renormalons. Solid lines denote quark propagators, while dashed lines are for gluons.
\label{bub}}
\end{figure}

Moving to higher orders in $\alpha$ we will find more and more complicated multiloop graphs. Among them a special role belongs to the bubble-chain diagrams, depicted in Fig. \ref{bub}.
These graphs (referred to as renormalons) were extensively studied in the late 1970s \cite{renormalon} (for reviews see \cite{beneke,ural}).

When we say bubble chains, we should be careful. Generally speaking, the very definition of a bubble chain in the form of Fig. \ref{bub} is not quite accurate. The appropriate renormalon graphs cannot be isolated in the form of a bubble chain since in this form they are not even gauge invariant. An honest-to-god renormalon calculation is quite contrived. 

There is a useful trick, however. One adds to the theory $N_f$ flavors, where $N_f$ is treated as a free parameter.
Then, instead of the full calculation of the genuine ``bubble chain," with gluon degrees of freedom in the bubbles, one calculates only the matter bubbles (which are gauge invariant in the chain of Fig. \ref{bub}), and then replaces
\beq
\beta_0^f\equiv -\frac 23 N_f \to \beta_0
\label{f31}
\eeq
where $\beta_0$ is the first coefficient in the $\beta$ function which includes everything: gluons (plus ghosts in the covariant gauges) and matter fields.\footnote{ Note that $\beta_0^f$ 
and $\beta_0$ have opposite signs.} 

It is easy to see that the renormalon contribution into the $D$ function is sigh-{\em non}alternating and factorially divergent in high orders,
$\Delta D_{\rm renorm} \sim n!\,\alpha^n$ ($n\gg 1$). If $n$ is large, the estimate $k\sim Q$ is no longer valid.
Both observations -- the absence of sign alteration and factorial divergence -- become obvious after a closer look at Fig. \ref{fig:w} {\em before integrating} over $k$. 
The exact result for fixed $k^2$ was found by Neubert \cite{MNe}. For  illustrative purposes it is sufficient
to use a simplified interpolating expression \cite{vzak} collecting all  bubble 
insertions
in the gluon propagator: no   bubbles, one   bubble, two   bubbles, and so on,
\beq
D= C \times Q^2 \int
dk^2\frac{k^2\alpha_s(k^2)}{(k^2+Q^2)^3} \,,
\label{renormp}
\eeq
which coincides with the exact expression \cite{MNe} in the limits
$k^2\ll Q^2$ and $k^2\gg Q^2$, up to minor irrelevant details.
The coefficient $C$ in Eq. (\ref{renormp}) is a numerical constant and $\alpha(k^2)$
is the running gauge coupling which can be represented as
\beq 
\alpha (k^2) = \frac{\alpha (Q^2)}{1-\frac{\beta_0\alpha (Q^2)\rule{0mm}{4mm}}{4\pi} 
\ln (Q^2/k^2)} \,.
\label{eightp}
\eeq

\begin{figure}[h]
   \epsfxsize=9cm
   \centerline{\epsffile{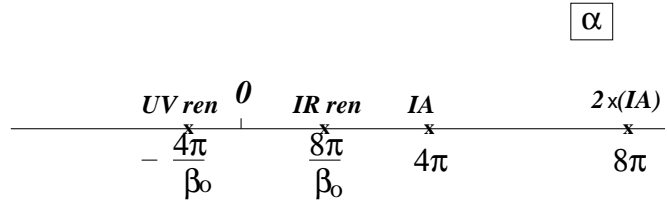}}
\caption{\small The Borel plane for the Adler function in QCD. The singularity to the left of the origin is due to an ultraviolet renormalon which need not concern us here. The nearest singularity to the right of the origin is due to the IR renormalon shown in  Fig. \ref{bub}. The IA singularities lie much further to the right.
\label{bpft}}
\end{figure}

Let us focus on the infrared (IR) domain. Omitting the overall constant $C$ we obtain 
  \beq
  D (Q^2)= \frac{1}{Q^4} \,\alpha \, \sum_{n=0}^\infty \left(\frac{\beta_0\alpha}{4\pi} 
  \right)^n  \int\,dk^2 \,k^2 \left(\ln \frac{Q^2}{k^2}
  \right)^n\,,\qquad
  \alpha \equiv \alpha(Q^2)
  \label{ninea}
  \eeq
  which can be rewritten as
   \beq
  D (Q^2)=  \frac{\alpha}{2}\,\sum_{n=0}^\infty  \left(\frac{\beta_0\alpha}{8\pi} 
  \right)^n  \int\,dy \,
    y^n\, e^{-y}\,,\qquad
 y = 2 \ln \frac{Q^2}{k^2}\,.
  \label{ninepppp}
  \eeq
The $y$ integration in Eq. (\ref{ninepppp}) represents all bubble-chain diagrams after integration over the loop momentum $k$.
The $y$ integral from zero to infinity is $n!$. A characteristic value of $k^2$
saturating the integral is
\beq
y\sim n   \,\,\,{\rm or}\,\,\,   k^2 \sim Q^2\,\exp \left(-\frac{n}{2}\right)  \,.
\label{tenp}
\eeq
Thus, we observe the factorial divergence of the
coefficients. The corresponding singularity in the Borel plane is depicted in Fig.  \ref{bpft}.

\begin{figure}[t]
   \epsfxsize=10cm
   \centerline{\epsffile{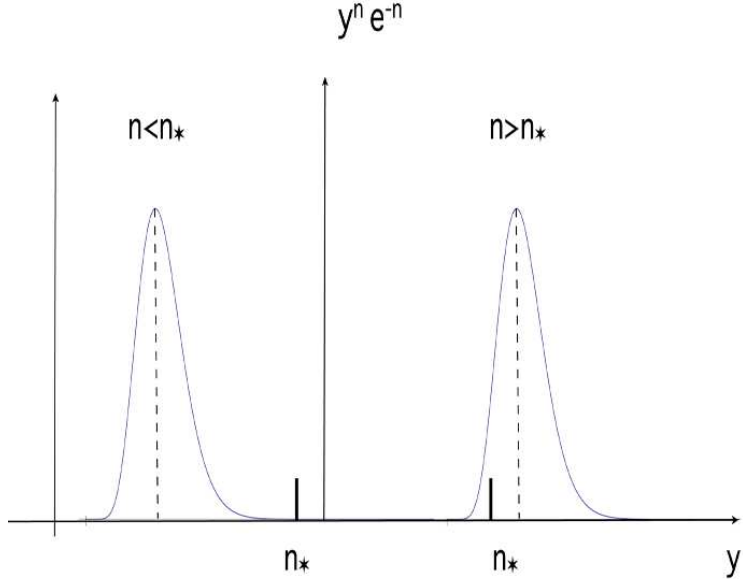}}
\caption{\small The plot of the integrand in Eq. (\ref{ninea}) for two values of $n$, ``small" 
and ``large." A sharp peak at $y\sim n$
saturates the integral. In the left plot $n<n_*= 2\ln(Q^2/\Lambda^2)$ and the forbidden 
domain $k^2\sim\Lambda^2$ does not contribute to the factorial factor.  In the right plot $n>n_*$. The $y$ integration has to be cut off at $y=n_*$,
which kills the factorial growth.
}
\label{peak}
\end{figure}

If $Q^2$ is fixed and $n$ is sufficiently large, $n>n_*$ where
\beq
n_* = 2\ln\frac{Q^2}{\Lambda^2}\,,
\label{r2}
\eeq
the factorial divergence of 
the coefficients in (\ref{ninepppp}) is purely formal and cannot be trusted.
At small $k^2\lsim \Lambda^2$ the diagrams in Figs.~\ref{fig:w} and \ref{bub} (in fact, any Feynman diagrams)
cease to properly 
represent non-Abelian dynamics  due to strong coupling in 
the IR. Equation (\ref{tenp}) shows that if $n>n_*$ the characteristic values of $k^2$
saturating the  integral do  fall off below $\Lambda^2$.
The point $n=n_*$ represents the optimal truncation point: at this point the terms of the asymptotic series are minimal. 
Formally, if we discard the  domain $k^2 < \Lambda^2$  at $n>n_*$ the factorial growth is suppressed, see Fig. \ref{peak},
and the series must be truncated,
\beq
 D (Q^2)\to   \frac{\alpha_s}{2}\,\sum_{n=0}^{n_*} \left(\frac{\beta_0\alpha_s}{8\pi} \right)^n  n! \,.
 \label{r2}
\eeq

At this point the road we have to take in QCD and similar strongly coupled theories diverges from that
in quantum mechanics. In the latter the validity of the quasiclassical approximation combined with the clear-cut picture of the vacuum structure allows one to achieve full resurgence.
In field theories the vacuum structure is determined by   infrared dynamics the  theory of which is still lacking, and quasiclassical approximations are bound to fail.  What can one do under the circumstances?

\section{Operator product expansion}

I remember that after the first seminar on the SVZ sum rules \cite{svz} in 1978 Eugene 
Bogomol'nyi used to ask me each time we met:
``Look, how can you speak of power corrections in the two-point functions at large $Q^2$ 
when even the perturbative expansion (i.e. the expansion in $1/\ln(Q^2/\Lambda^2)$) is 
not well defined? Isn't it inconsistent?" 

Now, with the discussion of Sec. \ref{ots}  in mind, I will be able to answer the above Bogomoln'yi question in a positive way, namely: 

\newpage

\begin{center}
$\star\star\star$ 
\end{center}
Consistent use of Wilson's OPE makes
everything well-defined at the conceptual level. Technical implementation may not always 
be straightforward, however. Moreover, the resulting OPE formula contains unknown vacuum condensates
in the form of power corrections. In turn, their summation presents an unsolved problem. 
\begin{center}
$\star\star\star$ 
\end{center}
 
The operator product expansion in asymptotically free theories is a {\em book-keeping device}
separating short-distance (weak-coupling) contributions from
those coming from large distances (strong coupling domain). To this end one introduces an auxiliary
an auxiliary 
parameter $\mu$, a separation scale between large and short distances. The latter contribution resides in the
OPE coefficient functions $C_i (Q,\mu)$, while the former contribution is encoded in the matrix elements
of the corresponding operators $O_i (\mu, \Lambda)$,
\beq
D(Q, \Lambda) = \sum_{i=0}^\infty \, C_i (Q,\mu)\, \left\langle O_i (\mu, \Lambda)
\right\rangle .
\label{38ya}
\eeq

 Generally  speaking,
 OPE is applicable whenever one deals with problems that can be formulated in the Euclidean 
space-time and in which
one can regulate typical Euclidean distances by a varying large external momentum $Q$.  
Factorization (\ref{38ya})  is technically meaningful (i.e. allows one to carry out
constructive calculations of $ C_i (Q,\mu)$) if one can choose
\beq
\mu\ll Q \,,\quad {\rm but}\,\,\, \mu\gg\Lambda
\label{39ya}
\eeq
Then the coefficients $ C_i (Q,\mu)$ can be found quasiclassically, even though they by no means reduce to PT.
The matrix elements $\left\langle O_i (\mu, \Lambda)
\right\rangle$ cannot be determined quasiclassically.
As a book-keeping device OPE cannot fail \cite{svz}, provided no arithmetic 
mistake is made {\em en route}. 

A remarkable observation was made in 1990s. 
Perturbative analysis (e.g. that of renormalons) prompts us that
certain nonperturbative condensates must be present in QCD. Moreover, one can even determine their dimension  
 from the  the position of singularities in the Borel plane. Unfortunately, by far not all condensates are visible in the analysis of PT  high orders. For instance, all condensates related to the spontaneous breaking of the chiral symmetry
leave no trace in any order of perturbation theory, nor in its divergence.

The values of condensates which are visible in the PT divergence are not determined by the PT analysis either.\footnote{ Prevalent in the 1970s and early 80s was a misconception that
the OPE coefficients are determined exclusively by perturbation theory
while the matrix elements of the operators involved are purely 
nonperturbative. Attempts to separate
perturbation theory from ``purely nonperturbative" condensates
gave rise to inconsistencies (see e.g.
\cite{David}). } 

\section{OPE and renormalons in QCD}

After this brief digression let us return to the Adler function  (\ref{currentcorr}) at large 
Euclidean $q^2$ in which OPE can be consistently 
built through separation of large- and short-distance contributions.

For simplicity, taking into account that my purpose today is 
illustrative, I will ignore the second inequality in (\ref{39ya}) and set the separation scale at $\mu=\Lambda$ rather than at $\mu\gg\Lambda$. 
This would be inappropriate in quantitative analyses; however, my task is to explain a qualitative situation.
Being auxiliary the parameter $\mu$ will  cancel from the master formula
(\ref{ope}) anyway, see  below.

\begin{figure}[h]
   \epsfxsize=9cm
   \centerline{\epsffile{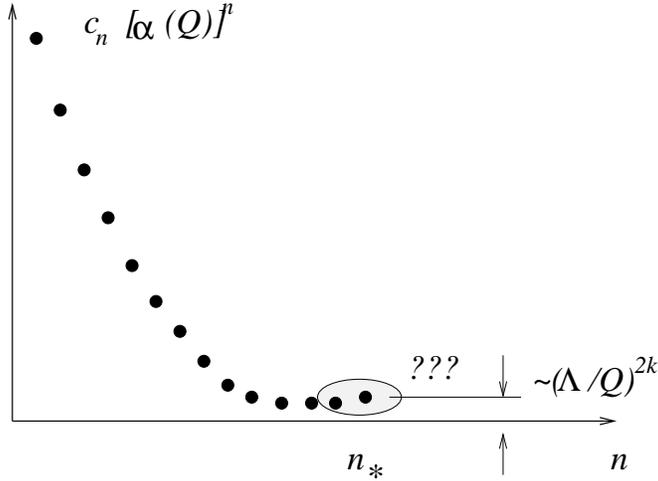}}
\caption{\small The PT expansion for  for the Adler function is asymptotic. One can trust it only up to a point of optimal truncation. A ``tail" beyond this point tells us of the existence of an operator of dimension $2k$ representing this tail
not accessible by PT calculation. (In the case at hand $k=2$).}
\label{asss}
\end{figure}

Let us have a closer look at Eqs. (\ref{renormp}) and (\ref{eightp}). The unlimited factorial 
divergence in (\ref{ninepppp}) is a direct consequence of integration over $k^2$ in (\ref{ninea}) 
all the way down to $k^2=0.$ Not only this is nonsensical because
of the pole in (\ref{eightp}) at $k^2=\Lambda^2$, this is {\em not} what we should  do in 
calculating coefficient functions in OPE. The coefficients must include $k^2>\Lambda^2$ by construction. The domain of small
$k^2$ (below $\Lambda^2$) must be excluded from $c_0$ and referred to the vacuum matrix 
element of the gluon operator
$G_{\mu\nu}^2$.
Indeed, in the sum in Eq. (\ref{ninea}) all terms with $n>n_*$ can be written as (see Figs. \ref{peak} and \ref{asss})
   \beqn
 \Delta D (Q^2)&=&  \frac{\alpha}{2}\,\sum_{n>n_*}  \left(\frac{\beta_0\alpha}{8\pi} \right)^n  n_*^ne^{-n_*}  
 \nonumber\\[3mm]
 &=&\frac{\alpha}{2}\,\sum_{n>n_*}  \frac{\Lambda^4}{Q^4}
 \label{35p}
  \eeqn
  where I used the fact that $$\frac{\beta_0\alpha(Q^2)}{8\pi}  =\frac{1}{2\ln(Q^2/\Lambda^2)}= 1/n_*\,.$$
  Of course, we can{\em not} calculate the gluon condensate from the above expression for the tail of the series
  (\ref{ninea}) representing the large distance contribution, for a number of  reasons. In particular, 
  the value of the coefficient in front of 
  ${\Lambda^4}/{Q^4}$ remains uncertain in (\ref{35p})  because Eq. (\ref{eightp}) is no longer 
  valid at such momenta. 
  We do not expect the gluon Green functions used in calculation in Fig. \ref{bub} and in 
  Eq. (\ref{eightp}) to retain any meaning in the strong coupling nonperturbative domain. 
  A qualitative feature -- the power dependence 
  $(\Lambda /Q)^4$ in (\ref{35p})
 -- is correct, however.
  
We note with satisfaction that the fourth 
  power of the parameter $\Lambda/Q$ which we find from this tail exactly matches the OPE contribution of
the operator  $\langle G_{\mu\nu}^2\rangle$, see \cite{svz}.

Summarizing, we see that the consistent use of OPE cures the problem of the 
renormalon-related factorial divergence of the coefficients
in the $\alpha$ series, absorbing the IR tail of the series in the vacuum expectation value 
of the gluon operator
$G_{\mu\nu}^2$ and similar higher-order operators. Although the value of $\langle G_{\mu\nu}^2\rangle$ 
cannot be calculated from renormalons, the
very fact of its existence can be established.

\section{Sources of factorials and master formula }

From quantum mechanics we learned that the factorial divergence can arise from  instantons.
In QCD the instantons are ill-defined in the IR and, strictly speaking, 
nobody knows what to do with them.\footnote{This statement is a slight exaggeration. 
I refer to \cite{shuryak}
for an alternative point of view on instantons in the QCD vacuum.} If one considers QCD in the 't Hooft limit of large number of 
colors instantons decouple. The corresponding singularity in the Borel plane (see Fig. \ref{bpft})
moves to the right infinity. At the same time, none of the essential features of 
QCD disappears in the 't Hooft limit.  Therefore, in our simplified consideration we can forget about instantons.
Perhaps, they will be needed later.

If so, we can write down  a single (simplified) ``master" formula for QCD and similar theories. 
At large Euclidean momenta the correlation functions of the type (\ref{currentcorr}) 
can be represented as
\beqn
D(Q^2)
&=&
\sum_{n=0}^{n_*^0}c_{0,n}\left(\frac{1}{\ln Q^2/\Lambda^2}\right)^n
\nonumber\\[3mm]
&+&
\sum_{n=0}^{n_*^1}c_{1,n}\left(\frac{1}{\ln Q^2/\Lambda^2}\right)^n\left(\frac{\Lambda}{Q}\right)^{d_1}
\nonumber\\[3mm]
&+&
\sum_{n=0}^{n_*^2}c_{2,n}\left(\frac{1}{\ln Q^2/\Lambda^2}\right)^n\left(\frac{\Lambda}{Q}\right)^{d_2}+ ...
\nonumber\\[3mm]
&+&
\mbox{``exponential terms"}\,.
\label{ope}
\eeqn
Equation (\ref{ope}) is simplified in a number of ways. First, it is assumed that
the currents in the left-hand side have no anomalous dimensions, and so do the operators 
appearing on the right-hand side. They are assumed to have only normal dimensions given 
by $d_i$ for the $i$-th operator. Second, I ignore the second and all higher coefficients 
in the $\beta$ function so that the running coupling is represented by a pure logarithm.
All these assumptions are not realistic in QCD.\footnote{They could be made somewhat more realistic in
${\mathcal N}=2$ super-Yang-Mills.} I stick to them to make the master formula more concise. 
Inclusion of higher orders in the $\beta$ function and anomalous dimensions both on the 
left- and right-hand sides will give rise
to rather contrived additional terms and factors containing $\log\log$'s,  $\log\log\log$'s 
$(\log\log/\log)$'s, etc.\footnote{Multiple logarithms are elements of the trans-series analysis too, see
\cite{edg}.} 
This is a purely technical, rather than conceptual,  complication, however. 

So far I discussed the divergence/convergence of the perturbative series 
explaining that the regulating parameter $\mu$
in OPE allows one to make PT meaningful.\footnote{Factorial divergence of PT series due to 
a factorially large number of  Feynman graphs with many loops is suppressed in the 't Hooft limit.} The expansion (\ref{ope}) runs not only in powers
of $1/\ln Q^2$, but also in powers of $\Lambda/Q$. This is a double expansion, and the 
power series in  $\Lambda/Q$ is also infinite in its turn. Does it have a finite radius of convergence?

Needless to say, this is an important question.  The answer to it  is {\em negative}.
Twenty years ago I argued in \cite{recent} (see also \cite{ope1}) that the
power series in (\ref{ope}) are factorially divergent in high orders. This is a rather straightforward
observation following  from the analytic structure of $D(Q^2)$. In a nutshell, since the $Q^2$ singularities 
in $D(Q^2)$ run all the way to infinity along the positive real semi-axis of $q^2$, the $1/Q^2$ expansion cannot be convergent. The last line in Eq. (\ref{ope}) symbolically represents a divergent tail of the power series. 

\section{Supersymmetric Yang-Mills}
\label{symtheo}

Factorial divergence of the perturbative series in supersymmetric theories was only scarcely discusses in the past \cite{Russo,z,y,R2}.  One can say that only first steps were made. Meanwhile, this is an interesting question because renormalons in supersymmetric theories 
have peculiarities related to peculiarities of the operator product expansion in supersymmetric Yang-Mills. 

As we already know, the renormalons  are in one-to-one correspondence with particular  gluon operators in OPE. There is a one-to-one correspondence between the given bubble chain graph and an appropriate operator in OPE (e.g. \cite{ural}).

The SYM Lagrangian is
\beq
{\mathcal L} 
=
 -\frac{1}{4g^2}G_{\mu\nu}^aG_{\mu\nu}^a
+\frac{i}{g^2}\bar\lambda^{a}\bar\sigma^\mu\, {\mathcal D}_{\mu} \lambda^{a}\,.
\label{syml}
\eeq
The only difference with the QCD Lagrangian (\ref{qcdl}) is in the fermion sector: the fundamental quarks are
replaced by a Majorana spinor in the adjoint representation of the gauge group.

Supersymmetry of  the model implies that an infinite  class of gluonic operators cannot have nonvanishing vacuum expectation values (VEVs). This fact tells us that the conventional renormalon
analysis must be modified.  Below I will discuss a modification needed, but at first
let us see why gluonic operator VEVs must vanish in SYM theory, in contradistinction to QCD.

\subsection{Why gluon operators have vanishing VEVs in SYM?}
\label{81}

The operator $G_{\mu\nu}^aG_{\mu\nu}^a
+ {i}\bar\lambda^{a}\bar\sigma^\mu\, {\mathcal D}_{\mu}
 \lambda^{a}
$
is the highest component of Tr$W^2$ where
\beq
W_\alpha = i\left(\lambda_\alpha + i\theta_\alpha\, D -\theta^\beta G_{\alpha\beta} - i \theta^2 \,
{\mathcal D}_{\alpha\dot\alpha}\bar\lambda^{\dot\alpha}\right).
\label{f81}
\eeq
Supersymmetry allows only the lowest
components of super fields to develop a nonvanishing VEV. 
In pure SYM theory, without matter, $D=0$ and, therefore, 
\beq
G_{\alpha\beta}\sim D_{\{\beta} W_{\alpha\}} + ...
\label{f10}
\eeq
where the braces denote symmetrization, $D_\beta$ is the spinorial derivative, and the ellipses stand for higher components irrelevant for our purposes. 

Gluonic operators in pure SYM theory must contain at least two $G$ factors; in other words their generic form is
\beq
O_g \propto G ... G \propto D_{\{\beta} W_{\alpha\}} ... D_{\{\tilde\beta} W_{\tilde\alpha\}}\,.
\label{11}
\eeq
The ellipses above represent any number of covariant derivatives and extra $W$ factors provided that
the overall number of the $W$ factors must be even. 
Taking Tr (which singles out color-singlet parts) is implied but not explicitly indicated.

The right-hand side in (\ref{11}) can be identically rewritten as 
\beq
D_{\{\beta} W_{\alpha\}} ... D_{\{\tilde\beta} W_{\tilde\alpha\}} = D_{\{\beta} \left( W_{\alpha\}} ... D_{\{\tilde\beta} W_{\tilde\alpha\}}\right) +  W_{\alpha }\left( D_{\beta}... D_{\{\tilde\beta} W_{\tilde\alpha\}}\right)\,.
\eeq
The first term is a full superderivative and, as such, can have no nontrivial VEV.
The lowest component of second term contains at the very least $\lambda$ and ${\mathcal D}_{\alpha\dot\alpha}\bar\lambda^{\dot\alpha}$ (the last factor vanishes due to the equation of motion).
Thus, the lowest-dimensional operator which could in principal appear in OPE is a two-$\lambda$ operator. However, this cannot appear too because if we calculate the OPE coefficients perturbatively (and renormalons are perturbative objects) then  two-$\lambda$ operators have wrong $R$ parity, while $\lambda\bar\lambda$ operators can have Lorentz spin zero only in the combination
\beq
{\rm Tr} \, \lambda_\alpha   {\mathcal D}_{\alpha\dot\gamma} G^{\dot\gamma\dot\beta} \bar\lambda^{\dot\beta}
\eeq
which reduces to a four-fermion operator by virtue of the equation of motion. 
Thus, in pure SYM we start OPE, in fact, from four-gluino operators
 of dimension 6, and the four-lambda operators with possible additional insertions of covariant derivatives and $G$ or $\lambda \lambda$ or $\lambda \bar\lambda$ factors  which have dimensions higher than 6. No purely gluonic operator can have a nonvanishing VEV in SYM. 
 
 The above argument based on $R$ parity is applicable to the two-point functions of the type
 \beq
 i\int \, d^4x\, e^{iqx}\left\langle O(x),\, O^\dagger(0)
\right\rangle\,,
\quad
O = {\rm Tr}\, \bar\lambda_{\dot\alpha} \lambda_\alpha \,\,\, {\rm or} \,\,\, {\rm Tr}\, \lambda^{\alpha} \lambda_\alpha\,,
\eeq
 
 Since the IR renormalons are in one-to-one correspondence with OPE, we conclude that the bubble chain in Fig. \ref{bub}
 which normally is responsible for non-Borel summable divergence of high orders (closest to the origin in the
 Borel plane) must be canceled by something else.

Let us note, however, that an easy way of identifying 
the ``bubble chains" in QCD was through matter loops with the subsequent extraction of the $N_f$ factor plus 
substitution (\ref{f31}). This trick does not work in SYM, see (\ref{syml}). We are forced to introduce matter fields.

\subsection{Matter loops in SYM}
\label{sec1}

To identify the renormalon bubble chain through matter loops  we
should expand supersymmetric gluodynamics (\ref{syml}) to include $N_f$ matter fields in
the fundamental representation of SU$(N)$,
\begin{eqnarray}
{\mathcal L} 
&=&
 -\frac{1}{4g^2}G_{\mu\nu}^aG_{\mu\nu}^a
+\frac{i}{g^2}\bar\lambda^{a}\bar\sigma^\mu\, {\mathcal D}_{\mu}
 \lambda^{a}
 \nonumber\\[3mm]
 &+& \sum_f\left(
 {\mathcal D}^\mu\overline{q_{\! f\rule{0mm}{3mm}}}\,{\mathcal
D}_\mu
q_f +i \,\overline{\psi_{\! f\rule{0mm}{3mm}}}\,\bar\sigma^\mu\,{\mathcal D}_\mu \psi_f
\right)
 \nonumber\\[3mm]
 &+& 
  \left[-\frac{m}{2} \psi_\alpha^f\psi^\alpha_f + i\sqrt{2}
\left(\psi_f\lambda^a\,T^a \right)\,\overline{q_{\! f\rule{0mm}{3mm}}} +{\rm H.c.}\right] -V(q_f)\,,
\label{spinnn350}
\end{eqnarray}
where
\begin{equation}
V(q_f) = \frac{g^2}{2} \left(\sum_f  \overline{q_{\! f\rule{0mm}{3mm}}}
\,T^a\,q_f\right)^2 +\sum_f |m|^2\left| q_f\right|^2\,.
\label{spinnn351}
\end{equation}
Here $q$ and $\psi$ are the squark and quark fields, respectively. The   mass terms in 
(\ref{spinnn350}) and (\ref{spinnn351}) are irrelevant and can be safely omitted.\footnote{One should remember that
each flavor is represented by two squarks and two Weyl quarks, one in the fundamental and another in the
 antifundamental representation of SU$(N)$.}

\begin{figure}[h]
   \epsfxsize=5cm
   \centerline{\epsffile{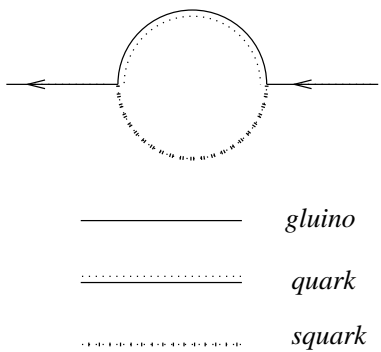}}
\caption{\small Elementary bubble insertion in the gluino line.
\label{fig4}}
\end{figure}

\begin{figure}[h]
   \epsfxsize=5cm
   \centerline{\epsffile{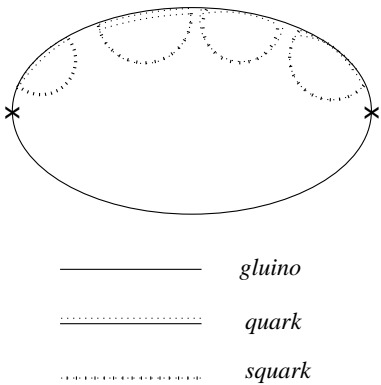}}
\caption{\small Additional bubble diagrams in SYM, with matter insertions in the gluino line. $N_f$ matter bubbles
produce the $N_f$ factor.
\label{fig3}}
\end{figure}

In addition to bubbles depicted in Fig. \ref{bub}, the bubble chains now develop elsewhere, e.g. Fig \ref{fig4}.
Unlike the familiar QCD example (Fig. \ref{bub}) in SYM theory the matter bubbles appear even in the diagrams without gluon insertions, such as the graph depicted in Fig. \ref{fig3}. Each bubble produces $N_f g^2 \log p^2 $ where $p$ is the momentum flowing through the gluino line.
However, this particular diagram would correspond to the operator 
\beq
 \bar\lambda_{\dot\alpha} i {\mathcal D}^{\dot\alpha \alpha}\lambda_\alpha
 \label{5}
 \eeq
  which reduces through the equation of motion to another dimension-4 operator,
namely, $$\sum_f \, \phi^j \left(\bar\lambda^i_j \bar\psi_i\right)\,,$$ which has no analog in QCD. Note that in SYM with matter chiral symmetry is broken {\em a priori}, and is replaced by $R$ symmetry of the U(1) type. 
In addition, there is an anomalous part in the equation of  motion to be used in Eq. (\ref{5}), which
produces the operator $G_{\mu\nu}{^2}$.

Calculation of the elementary bubble insertion in the gluino line is straightforward. It is determined by the 
graph in Fig. \ref{fig4}. In fact, there is no need for an explicit calculation of this diagram. 
It represents the $Z$ factor of the gluino field. However, supersymmetry guarantees that the renormalization of the gluon and gluino fields are identical. 

If we start from the Lagrangian (\ref{spinnn350}) normalized at a momentum scale $Q$ (then, the corresponding coupling is $g^2(Q)$) and evolve it down to $p$ then the operator $\frac{i}{g^2}\bar\lambda^{a}\bar\sigma^\mu\, {\mathcal D}_{\mu}
 \lambda^{a}$ in the Lagrangian
evolves in the following way
\beq
\frac{i}{g^2(Q)}\bar\lambda^{a}\bar\sigma^\mu\, {\mathcal D}_{\mu}
 \lambda^{a}
\to \frac{i}{g^2 (p) }\bar\lambda^{a}\bar\sigma^\mu\, {\mathcal D}_{\mu}
 \lambda^{a}\,.
 \label{6}
\eeq
The corresponding $Z$ factor can be easily read off, for instance, by proceeding to the canonically normalized gluino kinetic term. In this way we find
\beq
Z^{-1} = \frac{g^2(Q)}{g^2(p)} =1-\frac{g^2}{4\pi} \, (3N-N_f)\, \log\frac{Q^2}{p^2}\,,
\eeq
where the matter bubble produces only the $N_f$ part, of course. In other words, the (truncated) diagram in Fig. 
\ref{fig4} produces
\beq
\left(p_\mu \gamma^\mu \right)\frac{N_f g^2}{4\pi}\, \log\frac{Q^2}{p^2}\,.
\eeq

\vspace{2mm}

In summary, we see that the standard method of the renormalon  analysis which works well in QCD is not so straightforward 
in supersymmetric gluodynamics (i.e. gluons plus gluinos), since introduction of matter dramatically changes the OPE operator basis. In pure YM and in QCD with massless quarks it is one and the same dimension-4
 operator which acquires a VEV and is responsible for the leading renormalon singularity.
 This is in sharp contradistinction with what happens in SYM.

 \subsection{Renormalons, OPE, and diagrams in SYM}
 
\begin{figure}[h]
   \epsfxsize=5cm
   \centerline{\epsffile{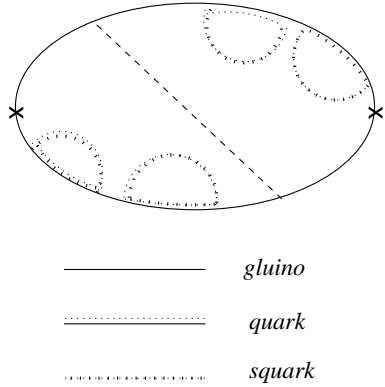}}
\caption{\small Lines with bubbles are soft. Those without bubbles are hard.
\label{fig5}}
\end{figure}

 Let us elucidate the last statement in more detail. 
The role of the IR renormalon bubble chain in a given diagram is to make the line to which bubbles are attached soft
\cite{beneke,ural}. At a critical value of $n$ the integration momentum flowing
 through the bubble  line becomes  of the order of $\Lambda$. Hence, in the framework of OPE, this line must be``cut" and becomes a part of the operator with a VEV, which will represent the tail of the renormalon. For instance, if we consider the graph in Fig. \ref{bub}, the solid lines carry large momentum, wile the dashed one is soft. Correspondingly,
 this bubble chain is in one-to-one correspondence with the operators $G^2$, $G {\mathcal D}^2G\to  G^3 $ and so on. In Fig. \ref{fig3} the upper part of the graph is soft, while the lower part is hard. One of the operators corresponding to this bubble chain is ${\rm Tr} \, \lambda_\alpha   {\mathcal D}_{\alpha\dot\beta}   \bar\lambda^{\dot\beta}$. Four-gluino operators   are obtained
 from the   chains depicted in Fig. \ref{fig5}.

In this graph one should ``cut" the gluino lines with the bubble insertions. A large external momentum is passed through the graph through the lines without bubbles. 
 
 The problem of interpretation arises only with the bubble chains attached to the gluon lines, because the corresponding operators which should conspire with the tail of such renormalons can have no VEVs. 
Basing on a arguments which I have no time to discuss  I am inclined to conjecture
 that the renormalon depicted in Fig. \ref{bub}
is canceled by the renormalon depicted
in Fig. \ref{fig3}. 
If the numerical coefficient $c$ is  right, the operator which these two graphs (combined together) will give in OPE 
\beq
 -\frac{1}{4}G_{\mu\nu}^aG_{\mu\nu}^a
+\frac{ic}{g^2}\,\bar\lambda^{a}\bar\sigma^\mu\, {\mathcal D}_{\mu}\,
 \lambda^{a} \to 0\,.
\eeq
This question should be explored further, however, see \cite{yi}.

\section{Conclusions}

1) Resurgence in the sense it is carried out in quantum mechanics, encounters with conceptual difficulties in strongly coupled Yang-Mills theories.

\vspace{2mm}

\noindent
2) The best we can do is to use Wilson's operator product expansion adapted to QCD, which has conceptual similarities with the resurgence program.

\vspace{2mm}

\noindent
3) Renormalons is a fancy way to say that at large distances perturbation theory requires matching
with nonperturbative strong coupling regime (``conspiracy"). Large distance dynamics is intricate and is yet unsolved. 
No details of the matching procedure are known, except that it must be consistent with OPE. It does not have to be universal  and may well differ in passing from QCD to supersymmetric Yang-Mills. 

\vspace{2mm}

\noindent
3) In SYM theories there are obvious problems with renormalons, not addressed in the past,
which are not yet solved. They seemingly defy a conventional (QCD-like) conspiracy between OPE and the tails of the renormalon contributions which was discussed in detail in my talk.
It is obvious that this question should become a focus of future studies.

\section*{Acknowledgments}

Useful discussion with A. Cherman, G. Dunne, M. \"Unsal, and A. Vainshtein  are gratefully acknowledged. 

This work  is supported in part by DOE grant DE-SC0011842.

\end{document}